\documentclass[aps,prl,twocolumn,superscriptaddress,showpacs]{revtex4-1} 
\usepackage{graphicx,amsmath,amssymb,ulem}
\usepackage{verbatim}
\usepackage{ragged2e}
\usepackage{color}
\usepackage{braket}

\begin{document}
\title{Modified Two-Photon Interference Achieved by the Manipulation of Entanglement}
\author{P.~R.~Sharapova}
\affiliation{Department of Physics and CeOPP, University of Paderborn, Warburger Stra\ss e 100, D-33098 Paderborn, Germany}
\author{K.~H.~Luo}
\affiliation{Department of Physics and CeOPP, University of Paderborn, Warburger Stra\ss e 100, D-33098 Paderborn, Germany}
\author{H.~Herrmann}
\affiliation{Department of Physics and CeOPP, University of Paderborn, Warburger Stra\ss e 100, D-33098 Paderborn, Germany}
\author{M.~Reichelt}
\affiliation{Department of Physics and CeOPP, University of Paderborn, Warburger Stra\ss e 100, D-33098 Paderborn, Germany}
\author{C.~Silberhorn}
\affiliation{Department of Physics and CeOPP, University of Paderborn, Warburger Stra\ss e 100, D-33098 Paderborn, Germany}
\author{T.~Meier}
\affiliation{Department of Physics and CeOPP, University of Paderborn, Warburger Stra\ss e 100, D-33098 Paderborn, Germany}

\begin{abstract}
In this theoretical study we demonstrate that entangled states are able to significantly extend the functionality of Hong-Ou-Mandel (HOM) interferometers.
By generating a coherent superposition of parametric-down-conversion photons and spatial entanglement in the input channel,
the coincidence probability measured at the output changes from a typical HOM-type dip (photon bunching) into much richer patterns
including an anti-bunching peak and rapid oscillation fringes with twice the optical frequency.
The considered system should be realizable on a single chip using current waveguide technology in the $LiNbO_3$ platform.
\end{abstract}
\maketitle

Quantum entanglement and hyperentanglement are highly important, fundamental, and indefeasible features of quantum mechanics.
Entangled states have numerous applications in emerging technologies such as quantum computing \cite{DiVicenzo},
quantum cryptography \cite{Barrett, Groeblacher}, quantum teleportation \cite{Bouwmeester, Takesu, Kurtsiefer}, and quantum algorithms \cite{Bergou, Bourennane}.
The highly attractive hyperentangled states extend possibilities of quantum technologies, allowing new quantum communication protocols, superdense teleportation \cite{Graham}, and
objects with higher dimensionality \cite{Barreiro}.
Therefore the generation and application of photon states with a high degree of entanglement and hyperentangled states
is an intensely studied topic in quantum optics.

An important element in a computational infrastructure based on photons are quantum interferometers which allow to measure the degree of indistinguishability of photons.
Such interferometers can be realized using the well-known two-photon Hong-Ou-Mandel (HOM) interference, which was demonstrated
in several systems \cite{Bentivegna,Hong,Ou,Abouraddy,Kim,arxiv, Faruque}.

For practical large-scale applications in quantum information processing free-space set-ups are not very reliable
because of the experimental complexity that is required to achieve and maintain a precise and stable adjustment between the elements and due to the significant size of such systems.
However, due to their small size and high stability integrated quantum optical systems \cite{Pollock} are very promising in this direction.
Furthermore, they provide an attractive platform for the realization of a wide range of functionalities including quantum simulations \cite{AspuruGuzik, Georgescu},
boson sampling \cite{Tillmann, Crespi, Huh}, quantum computation, and quantum communication processing \cite{Nielsen}.

\begin{figure}[htb]
\begin{center}
\includegraphics[width=0.8\columnwidth]{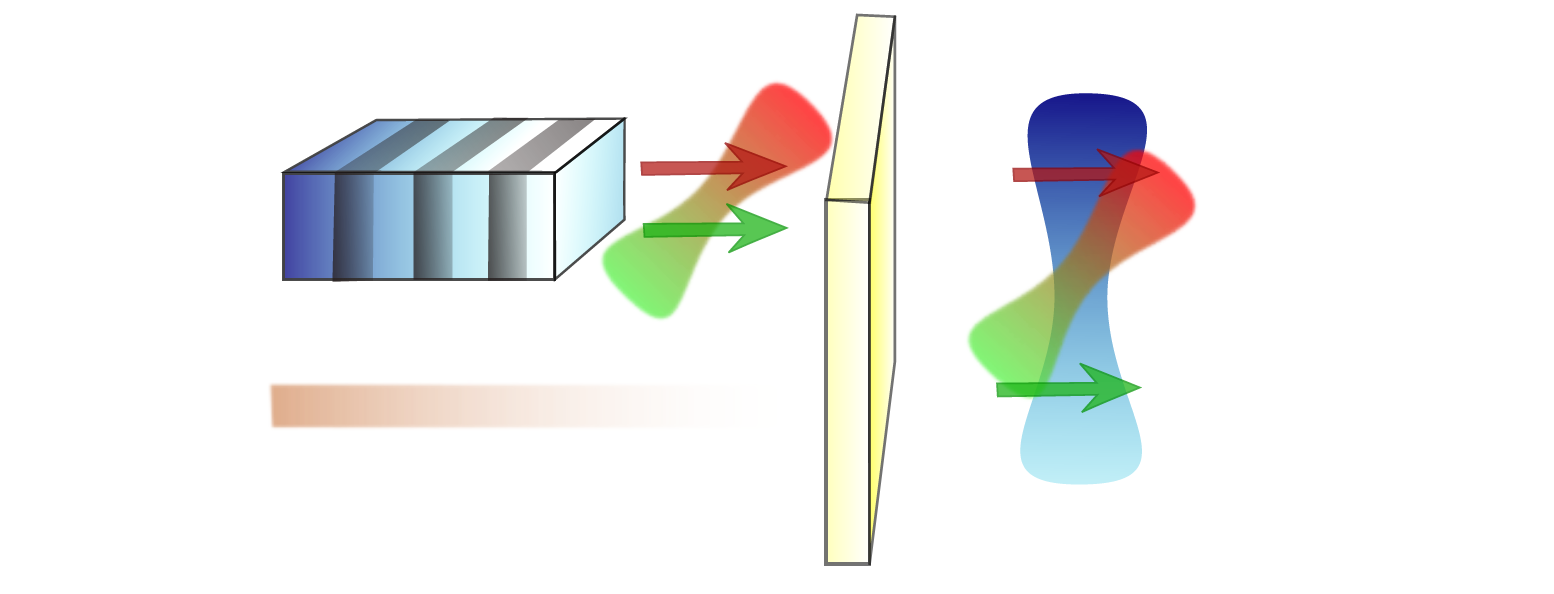}
\end{center}
\caption{Two photons generated in a PDC process have frequency correlations.
The creation of spatial entanglement between the photons leads to a hyperentangled structure
which results in novel features in the coincidence probability.}
\label{fig:idea}
\end{figure}

In this paper, we propose and analyze a device which can be realized on a single integrated platform and is able to manipulate the two-photon coincidence probability
by interference and to create hyperentanglement.
The photon pair generation is incorporated into the system using an on-chip parametric-down-conversion (PDC) \cite{Boyd, Louisell, Mollow, Mosley} source.
We theoretically show that the creation of spatially-entangled photons in a singlet Bell state is possible in such a device and that the entanglement leads to
novel features in the coincidence probability.
In particular, different regimes of bunching (HOM-type dips) and anti-bunching (singlet Bell state) and rapid interference fringes corresponding to twice the optical
frequency are predicted. 
Simultaneously, frequency correlations are preserved in the system which leads to the creation a hyperentanglement, see Fig.~\ref{fig:idea}.

\begin{figure}[htb]
\begin{center}
\includegraphics[width=1.0\columnwidth]{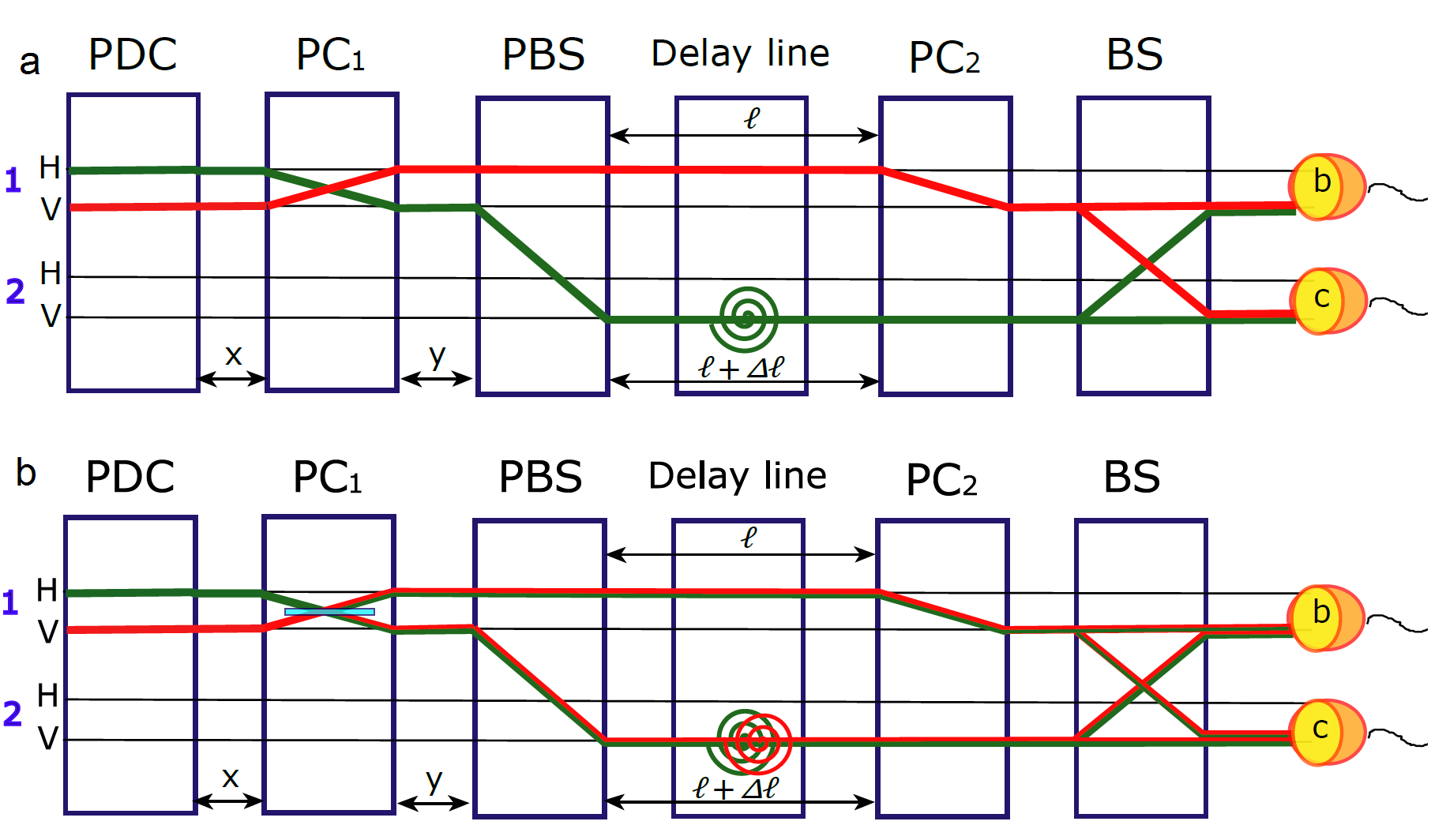}
\end{center}
\caption{Schematical illustration of the considered setup.
The propagation of the signal (idler) photon is depicted by the green (upper) (red (lower)) line.
Two photons with orthogonal polarizations H and V are created in the PDC section in the upper spatial channel 1.
The lower channel 2 is initially empty.
The first polarization converter $PC_1$ is placed in channel 1 and changes the polarizations by $\phi_1$.
The polarization-sensitive beam splitter $PBS$ spatially separates photons with orthogonal polarizations.
The delay section compensates the time delay between photons, $l$ is the length of the upper channel, $L=l+\Delta l$ is the length of the lower channel.
$PC_2$ changes the polarization in the upper channel 1 by $\phi_2$.
The coincidence probability is measured by the detectors $b$ and $c$.
(a) $PC_1$ converts the polarization perfectly, i.e., $\phi_1=\pi/2$.
(b) $PC_1$ partially converts the polarization, e.g., $\phi_1=\pi/4$ which can be realized by adding a beam splitter inside $PC_1$.}
\label{fig:sketch1}
\end{figure}

The photon pairs are created in an externally pumped type II degenerate PDC section realized by a periodically-poled Titanium-indiffused $LiNbO_3$ waveguide.
The generated signal and idler photons have orthogonal polarizations and therefore different group velocities.
In lowest-order perturbation theory the biphoton state created in the PDC section is described by \cite{Law}
\begin{equation}
\Ket{\psi_{PDC}}=\int d \omega_s d \omega_i F(\omega_s, \omega_i)a^\dagger_{1H}(\omega_s)a^\dagger_{1V}(\omega_i)\Ket{0},
\label{input}
\end{equation}
where $a^\dagger_{1H}(\omega_s)$ ($a^\dagger_{1V}(\omega_i)$) is the creation operator for the horizontally (H) (vertically (V)) polarized signal (idler) frequency mode,
the index $1$ denotes the upper channel of the system, see Fig.~\ref{fig:sketch1}, 
and $F(\omega_s, \omega_i)$ is the two-photon amplitude (TPA) \cite{TPA}.

To be able to observe HOM interference of the two PDC photons, basically two goals must be achieved
to ensure that indistinguishable photons arrive at the beam splitter ($BS$).
Firstly, the polarization of one of the photons has to be changed and, secondly,
the time delay between the two photons arising from the polarization-dependent group velocities needs to be compensated for.
For the realization of these goals and also for the on-chip generation of entanglement we suggest the setup sketched in Fig.~\ref{fig:sketch1}.
Two PDC photons with orthogonal polarizations are generated in the upper channel 1 and arrive at a first polarization converter ($PC_1$).
When $PC_1$ works perfectly, i.e., it is set to a rotation angle of $\phi_1=\pi/2$ \cite{pc},
the polarization of the photons is interchanged, see Fig.~\ref{fig:sketch1}(a), but they remain in channel 1.
By a polarization-sensitive beam splitter ($PBS$) the two photons are separated in two different spatial channels 1 and 2.
Then the polarization of the photon in the upper channel 1 is rotated by a polarization converter ($PC_2$) set to $\phi_2=\pi/2$ and simultaneously the length difference between
the upper and lower channels is chosen in order to compensate the time delay using a delay line in channel 2.
As a result, two indistinguishable photons will arrive at the final $BS$ allowing to observe HOM interference. The all mentioned elements can be realized on chip with using available current integrated technologies \cite{arxiv}.
When $PC_1$ is switched-off, i.e., $\phi_1=0$, the polarizations are initially not converted and thus the time delay between the signal and idler photons is changed in comparison
to the case $\phi_1=\pi/2$.
Also in this case a HOM dip arises, however, at a shifted temporal position.

The possibility to vary the rotation angle $\phi_1$ of $PC_1$ significantly increases the functionality of the considered set-up.
If $\phi_1$ is not equal to a multiple of $\pi/2$, see Fig.~\ref{fig:sketch1}(b), 
$PC_1$ puts both the signal and idler photons in coherent superposition of the two polarizations.
Afterwards, the $PBS$ separates photons with different polarizations into different channels.
Thus, the combined action of $PC_1$ and $PBS$ allows to create spatial entanglement between PDC photons and,
in fact, generates a singlet Bell state arriving at the final $BS$.

To theoretically evaluate the propagation of the PDC-generated photon pair through the set-up
we describe the evolution, i.e., the action of each element and the free propagations in between, by unitary matrices \cite{arxiv, parameters}.
The unitary evolution matrix of the entire system is given by a product of several matrices and reads
\begin{equation}
U_{total}=BS*FP_3*PC_2*FP_2*PBS* FP_1 * PC_1 *FP_0 ,
\label{transformation2}
\end{equation}
where the matrix $PC_1$ describes the polarization rotation by the angle $\phi_1$ in channel 1,
and $BS$, $PBS$, and $PC_2$ are the matrices of the beam splitter, the polarization-sensitive beam splitter, and the second polarization converter, respectively.
The diagonal matrices $FP_i$ include phase factors arising from the free propagations between the elements.
The coincidence probability of photons in the detectors $b$ and $c$ is defined as 
\begin{equation}
P_{b,c}^{\lambda, \lambda^{'}}=\int d \omega_b d \omega_c |\Bra{0} d_{1,\lambda}(\omega_b)d_{2,\lambda^{'}}(\omega_c) \Ket{\psi_{out}}|^2 ,
\label{POVM1}
\end{equation}
where $\Ket{\psi_{out}}$ is the state arriving at the detectors, $d_{1,\lambda}(\omega_b)$ and $d_{2,\lambda^{'}}(\omega_c)$ are annihilation operators at the detectors
for frequencies $\omega_b$ and $\omega_c$ and polarizations $\lambda$ and $\lambda^{'}$, respectively.

It is useful to evaluate that part of the wavefunction which describes a vertically-polarized photon pair
{\it before} its arrival at the $BS$. This is obtained by acting with 
\begin{equation}
U_{before\,\,BS} = FP_3*PC_2*FP_2*PBS* FP_1 * PC_1 *FP_0 
\label{transformation3}
\end{equation}
on the PDC state and then projecting onto vertical polarization for both photons.
Considering a perfect $PBS$ and $PC_2$ set to $\phi_2=\pi/2$,
the vertically-polarized photon pair arriving at the $BS$ depends on  $\phi_1$ and is given by
\begin{equation}
\Ket{\psi^{VV}_{out}}=\Ket{\psi_1}+\Ket{\psi_2},
\label{wfPi4}
\end{equation}
with 
 \begin{eqnarray}
\Ket{\psi_1}=i\int d \omega_s d \omega_i F(\omega_s, \omega_i) e^{i(\frac{\omega_s}{v_H}+\frac{\omega_i}{v_V})x}\times
\nonumber\\
 \lbrace\chi(\omega_s,\omega_i)\sin [\phi_1]^2 a^{\dagger}_{2V}(\omega_s)a^{\dagger}_{1V}(\omega_i)
\nonumber\\
 -\chi(\omega_i,\omega_s)\cos[\phi_1]^2 a^{\dagger}_{1V}(\omega_s)a^{\dagger}_{2V}(\omega_i)\rbrace
\label{wfPi4psi1}
 \end{eqnarray}
 and
 \begin{eqnarray}
\Ket{\psi_2}=-\int d \omega_s d \omega_i F(\omega_s, \omega_i) e^{i(\frac{\omega_s}{v_H}+\frac{\omega_i}{v_V})x}\times
\nonumber\\
 \sin[\phi_1] \cos[\phi_1] \lbrace\Phi_1(\omega_s, \omega_i)a^{\dagger}_{1V}(\omega_s)a^{\dagger}_{1V}(\omega_i) + 
\nonumber\\ 
 \Phi_2(\omega_s, \omega_i)e^{i(\omega_s+\omega_i)\frac{\Delta l}{v_V}} a^{\dagger}_{2V}(\omega_s)a^{\dagger}_{2V}(\omega_i)\rbrace, 
\label{wfPi4psi2}
 \end{eqnarray}
with $\chi(\omega_s,\omega_i)=\exp[i(\frac{l+\Delta l+y}{v_V}\omega_s+\frac{l+y}{v_H}\omega_i)]$, 
$\Phi_1(\omega_s, \omega_i)=\exp[i\frac{l+y}{v_H}(\omega_s+\omega_i)]$,
$\Phi_2(\omega_s, \omega_i)=\exp[i\frac{l+y}{v_V}(\omega_s+\omega_i)]$, $v_{H(V)}=c/n_{H(V)}$,
and $n_{H(V)}$ the refractive index for horizontally (vertically)-polarized light.

\begin{figure}[htb]
\begin{center}
\includegraphics[width=1.0\columnwidth]{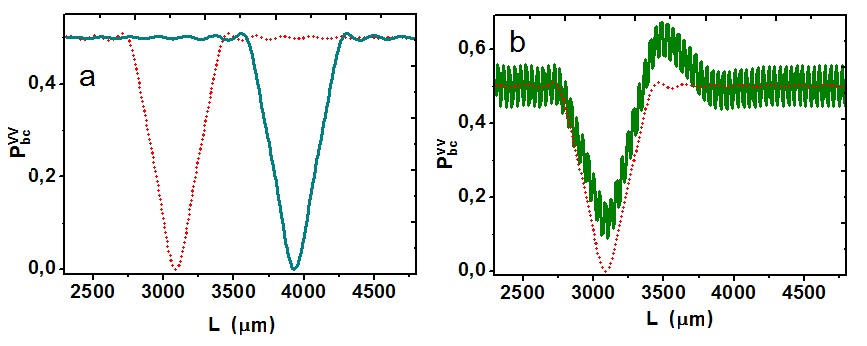}
\end{center}
\caption{The coincidence probability for two vertically-polarized photons for (a) different angles of $PC_1$: $\phi_1=0$ (cyan solid)
and $\phi_1=\pi/2$ (red dotted) and (b) considering $\phi_1=3 \pi/8$ (green solid)
in comparison with the case of $\phi_1=\pi/2$ (red dotted).} 
\label{fig:HOMfai}
\end{figure}

The density matrix of this state reduced over one spatial channel and over frequencies can be written as
\begin{equation}
\rho_r= \int d \omega_s d \omega_i |F(\omega_s, \omega_i)|^2 \begin{pmatrix}
\sin^2 \phi_1 & A(\omega_s, \omega_i, \phi_1) \\
A^{*}(\omega_s, \omega_i, \phi_1) & \cos^2 \phi_1 \\
 \end{pmatrix},
\label{reduced}
\end{equation}
with\\ 
$A(\omega_s, \omega_i, \phi_1)=\frac{i}{4}\sin 4\phi_1 \exp[i \omega_s [\frac{\Delta l}{v_V}-(l+y)(\frac{1}{v_H}-\frac{1}{v_V})]]$.
This expression shows that the reduced density matrix represents a pure state if $\phi_1$ is a multiple of $\pi/2$ 
For $\phi_1=\pi/4$ the off-diagonal elements vanish and the reduced density matrix describes an entangled state.
For other angles $\phi_1$ the off-diagonal elements are finite and the degree of entanglement depends also on the frequency components.

For $\phi_1=0$ $PC_1$ does not change the polarization of the photon pair.
So in this case a HOM dip is expected in the coincidence probability since according to Eq.~(\ref{wfPi4}) the two-photon wave function arriving at the $BS$ is
proportional to $a^{\dagger}_{1V}(\omega_s)a^{\dagger}_{2V}(\omega_i)$.
This is confirmed by the cyan solid curve in Fig.~\ref{fig:HOMfai}(a) showing the coincidence probability versus the length $L$ of channel 2.
Setting $PC_1$ to $\phi_1=\pi/2$ flips the polarizations of both photons.
This reduces the time delay of the photons corresponding to a smaller length
of the second channel at which the HOM dip appears, see red dotted curve in Fig. ~\ref{fig:HOMfai}(a).
So again we find, as expected, an ordinary HOM dip as the two-photon wave function arriving at the $BS$ is
proportional to $a^{\dagger}_{2V}(\omega_s)a^{\dagger}_{1V}(\omega_i)$, i.e., compared to $\phi_1=0$ only the channel indices are interchanged.
However, when $\phi_1$ is not a multiple of $\pi/2$ the situation changes significantly since the vertically-polarized photons arriving at the $BS$ are
according to Eq.~(\ref{wfPi4}) described by a more complex wave function.
For example, when we consider $\phi_1=3 \pi/8$ the coincidence probability changes qualitatively.
As shown by the green solid line in Fig.~\ref{fig:HOMfai}(b), in this case besides a HOM-type dip of reduced size
also a peak appears and in addition rapid oscillations show up in the coincidence probability.
The physical origin of these features is analyzed further below.

\begin{figure}[htb]
\begin{center}
\includegraphics[width=1.1\columnwidth]{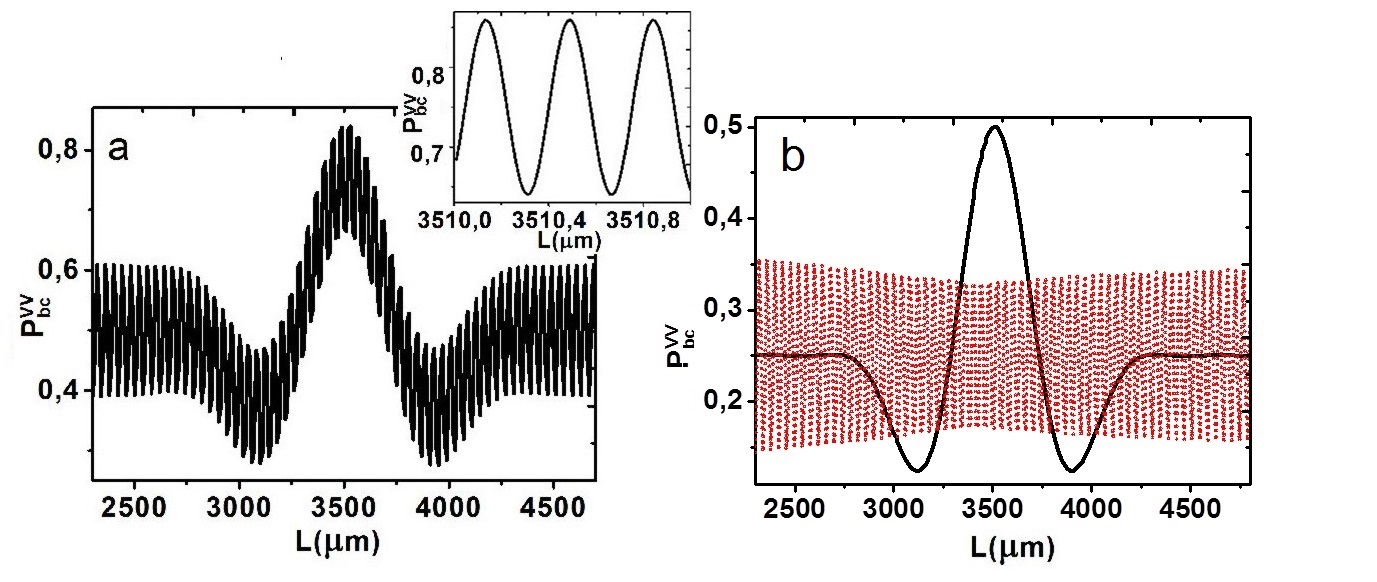}
\end{center}
\caption{(a) The coincidence probability for two vertically-polarized photons for $\phi_1=\pi/4$.
The inset shows a zoom-in on the rapid oscillations.
(b) The individual contributions of the two parts of the two-photon wavefunction to the coincidence probability:
Eq.~(\ref{wfPi4psi1}) (black solid) and Eq.~(\ref{wfPi4psi2}) (red dotted).}
\label{fig:HOMpi4it}
\end{figure}

The most interesting situation arises when the rotation angle of $PC_1$ is set to $\phi_1=\pi/4$.
In this case, all terms in $\Ket{\psi_1}$ and $\Ket{\psi_2}$ have the same prefactor and therefore
contribute with the same strength to the wave function
since $PC_1$ converts both the signal and the idler photons into a coherent superposition state of both polarizations with equal amplitudes.
Due to the polarization-dependent refractive index,
the propagation through the system leads to a finite time delay between the orthogonal polarization components of each photon.
If this time delay is compensated by a proper length of channel 2, in addition to the two photon interference,
now different parts of the same photon arrive at the $BS$ simultaneously.
Thus for $\phi_1=\pi/4$, Eq.~(\ref{wfPi4psi1}) represents a pure singlet Bell state and includes
the highest amount of entanglement between the signal and idler photons in the considered system.

The degree of entanglement can be characterized by the Schmidt number $K=1/Tr(\rho_r^2)$, where $\rho_r$ is the reduced density matrix, see Eq.~(\ref{reduced}).
For an appropriate length difference $\Delta l=(\frac{v_V}{v_H}|_{\frac{\omega_p}{2}}-1)(l+y)$ between the lower (length $L$) and upper (length $l$) channels 2 and 1,
the non-diagonal matrix element is
$A(\omega_s, \omega_i, \phi_1)=\frac{i}{4} \sin 4\phi_1$ and the Schmidt number
\begin{equation}
K=\frac{1}{\cos^{4}(\phi_1)+\sin^{4}(\phi_1)+\frac{1}{8} \sin^2 4 \phi_1}.
\label{Schmidt}
\end{equation}
Clearly, for both the non-converted situation and the converted case, i.e., $\phi_1=0$ or $\phi_1=\pi/2$,
the Schmidt number is $K=1$ which corresponds to a pure state before $BS$.
For any other value of $\phi_1$ the Schmidt number $K$ is larger than $1$ and correspondingly spatial entanglement is present.
This entanglement is maximal for $\phi_1=\pi/4$ where the Schmidt number is $K=2$.
The presence of entanglement leads to the antibunching peak in the coincidence probability visible in Fig.~\ref{fig:HOMpi4it}
that in the case $\phi_1=\pi/4$ is directly connected with the singlet Bell state arriving at the analyzer, i.e., the $BS$.

The positions of the two dips in Fig.~\ref{fig:HOMfai}(a) directly correspond to the positions of the HOM dips in Fig.~\ref{fig:HOMpi4it}.

In this case, however, the minima of the coincidence probabilities are significantly larger than zero because
besides indistinguishable polarization-converted photons at the same time also 
distinguishable by time non-converted by $PC_1$ photons arrive at the $BS$.
These two HOM-type dips appear for length differences between the lower and the upper channels of
$\Delta l=(\frac{v_V}{v_H}|_{\frac{\omega_p}{2}}-1)(l+y+x+\frac{L_{PDC}}{2})$ and $\Delta l=(\frac{v_V}{v_H}|_{\frac{\omega_p}{2}}-1)(l+y-(x+\frac{L_{PDC}}{2}))$, respectively.
This interpretation is confirmed by the black solid line in Fig.~\ref{fig:HOMpi4it}(b) which shows the contribution of Eq.~(\ref{wfPi4psi1}) alone to the total coincidence probability.

The second part of the wavefunction, Eq.~(\ref{wfPi4psi2}), corresponds to the situation that two photons are in the same channel when they reach the $BS$.
When the time delay between the two photons is sufficiently small such that the wave packets overlap, interference takes place.
In this case, rapid fringes corresponding to twice the optical frequency
which originate from the $(\omega_s+\omega_i)$-terms in Eq.~(\ref{wfPi4psi2}) are visible in the coincidence probability
manifesting the multimode structure of radiation in the frequency domain.
The frequency entanglement between signal and idler photons is preserved and depends on the number of modes in the frequency domain which for a long pump pulse can be very high.
Thus the creation of additional spatial entanglement can lead to large hyperentanglement.

The period of the rapid interference fringes can be estimated as
\begin{equation}
\omega_p \lbrace(\frac{1}{v_H}-\frac{1}{v_V})(l+y)-\frac{\Delta l}{v_V}\rbrace|_{\frac{\omega_p}{2}}=2\pi n
\end{equation}
which is approximately equal to $0.4 \mu m$ for the parameters used here, see inset of Fig.~\ref{fig:HOMpi4it}(a).
The red dotted curve in Fig.~\ref{fig:HOMpi4it}(b) shows the contribution of Eq.~(\ref{wfPi4psi2}) alone to the total coincidence probability presented in
Fig.~\ref{fig:HOMpi4it}(a).
Note that the sum of the two contributions shown in Fig.~\ref{fig:HOMpi4it}(b) is identical to the total result shown in Fig.~\ref{fig:HOMpi4it}(a).

\begin{figure}[htb]
\begin{center}
\includegraphics[width=1.05\columnwidth]{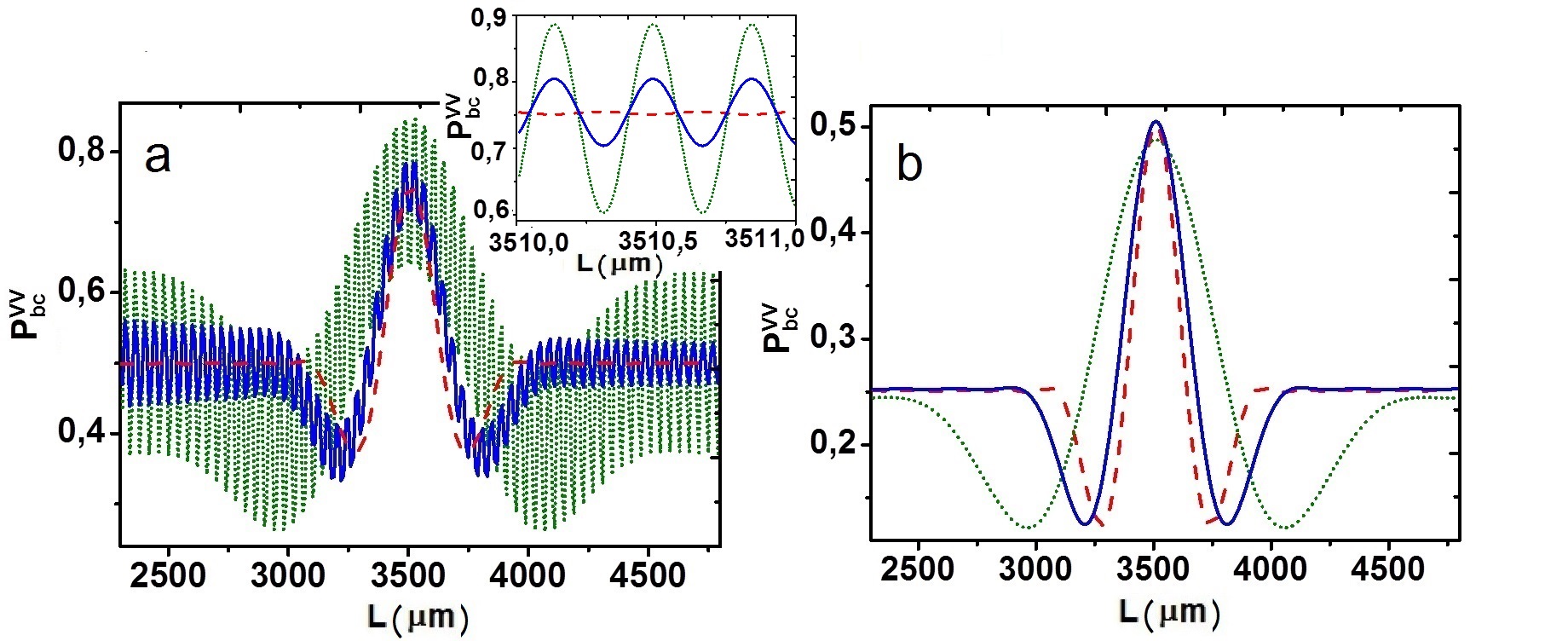}
\end{center}
\caption{The coincidence probability for $\phi_1=\pi/4$ for two vertically-polarized photons considering different PDC section lengths:
$L_{PDC}=1.035 cm$ (red dashed), $L_{PDC}=1.5 cm$ (blue solid), and $L_{PDC}=3.07 cm$ (green dotted).
(a) The total coincidence probability, where the inset zooms-in on the oscillations and
(b) the contribution of Eq.~(\ref{wfPi4psi1}) alone.} \label{fig:differentLPDC1}
\end{figure}

The amplitude of the rapid oscillations is directly related to the overlap of the two-photon wave packet.
It is possible to manipulate the magnitude of the oscillations by varying the length of the PDC section,
the length between the PDC section and $PC_1$ ($x$),
or the length between $PC_1$ and the $PBS$ ($y$).
Fig.~\ref{fig:differentLPDC1}(a) presents the coincidence probability for different PDC section lengths.
It is clearly visible that with increasing $L_{PDC}$ the amplitude of the oscillations grows because the wave packets spread more in the time domain and therefore their overlap region increases.
Fig.~\ref{fig:differentLPDC1}(b) shows the coincidence probability of only the contribution of Eq.~(\ref{wfPi4psi1}) to the two-photon wave function
and demonstrates the tendency of the increasing length difference and broadening of two HOM dips with increasing $L_{PDC}$.
When changing $L_{PDC}$ the position of the central peak remains unchanged because this depends only on $(y+l)$. 

\begin{figure}[htb]
\begin{center}
\includegraphics[width=1.0\columnwidth]{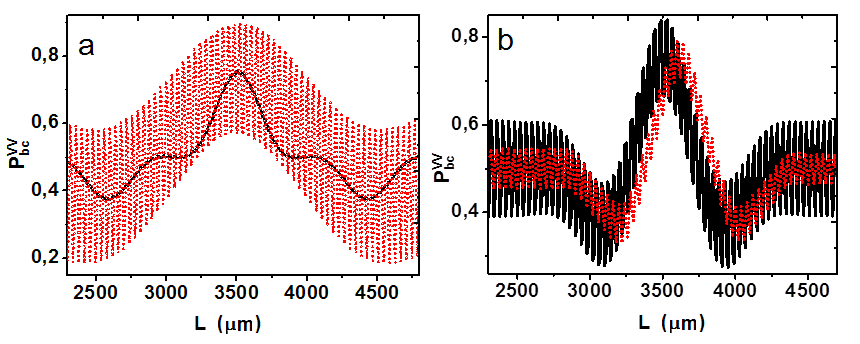}
\end{center}
\caption{The coincidence probability for two vertically-polarized photons for $\phi_1=\pi/4$ and:
(a) $L_{PDC}=2.07 cm$ and an additional length between the PDC section and $PC_1$ of $4.14 cm$,
$x=4.14 cm+L_{PC_1}/2=45210 \mu m$ (black solid)
and $L_{PDC}=6.21 cm$, $x=L_{PC_1}/2=3810 \mu m$ (red dotted) using $y=5810 \mu m$ for both cases.
(b) $y=5810 \mu m$ (black solid) and $y=9810 \mu m$ (red dotted) using $L_{PDC}=2.07 cm$ and $x=L_{PC_1}/2=3810 \mu m$ for both curves.}
\label{fig:differentxy}
\end{figure}

As mentioned above, it is possible to suppress the rapid oscillations by including additional distances between the PDC section and $PC_1$ or
between $PC_1$ and the $PBS$ for a fixed PDC section length. 
Fig.~\ref{fig:differentxy}(a) shows the coincidence probability for $L_{PDC}=2.07 cm$ and an added length of $4.14 cm$ between the PDC section and
$PC_1$ (black solid curve) and $L_{PDC}=6.21 cm$ (red dotted curve). 
In the first case, the wave packets of the two photons do not overlap any more and consequently the oscillations disappear.
The oscillations can also be suppressed by increasing the length between $PC_1$ and the $PBS$ ($y$).
In this case, the additional length increases the delay between the wave packets which decreases their overlap and accordingly the oscillation amplitude,
see Fig.~\ref{fig:differentxy}(b).
Simultaneously, shifting of the HOM dips and the central anti-bunching peak appears because their positions depend on $y$.
 
In conclusion, we demonstrate that the creation of spatially-entangled photons in a singlet Bell state is possible in the proposed integrated set-up
which contains a PDC photon-pair source and several optical elements. The proposed effects do not rely on the specific generation process for the photon pairs and thus also other sources could be used as long as the photonics properties are not modified.
Experimentally such a system should be realizable on a single chip using currently available $LiNbO_3$ waveguide technology.
The entanglement created in the system may lead to the appearance of an anti-bunching peak and rapid interference fringes with twice the optical frequency
in the two-photon coincidence probability.
By changing a free propagation lengths it is possible to manipulate the amplitude of these oscillations.
Simultaneously, frequency correlations between signal and idler photons present in the system lead to a hyperentangled structure that significantly extends the
capacity of the device.
The developed theoretical approach provides an intuitive understanding of the presented results and is also applicable to more complex set-ups.

Our findings open novel possibilities for modifying photonic states which can be useful for applications in emerging quantum technologies, e.g.,
quantum metrology, since the double frequency oscillations may allow very sensitive interferometry.
Furthermore, the realization of quantum logical operations, quantum communication and information processes, generation and easy manipulation by hyperentangled states
and Bell states generation promise significant novel functionalities for quantum optical devices.

\acknowledgments
Financial support of the Deutsche Forschungsgemeinschaft (DFG) through TRR~142, project C02, is gratefully acknowledged.
P.R.Sh. thanks the state of Nordrhein-Westfalen for support by the {\it Landesprogramm f\"ur geschlechtergerechte Hochschulen}.


\begin{thebibliography}{99}
\bibitem{TPA}For the pump laser we consider a Gaussian envelope, $E_p^{(+)} (\mathbf{r},t)= E_0 e^{-\frac{t^2}{2\tau^2}}e^{i( \mathbf{k}_p \cdot \mathbf{r}-\omega_p t)}$, with a full width at half maximum (FWHM) of the intensity of $2\sqrt{\ln 2}\tau$ where $\tau$ is the pulse duration.
The two-photon amplitude in this case is given by
\begin{eqnarray}
F(\omega_s, \omega_i)=C \exp [\frac{(\omega_s+\omega_i-\omega_p)^2}{2\Omega^2}]\mathrm{Sinc}[\dfrac{\Delta k \, L_{PDC}}{2}]\times 
\nonumber\\
\exp[\dfrac{i\Delta k\, L_{PDC}}{2}], \ \ \ \
\nonumber
\label{TPA}
\end{eqnarray}
where $C$ is a normalization constant, $\Delta k =k_p(\omega_s+\omega_i)-k_s(\omega_s)-k_i(\omega_i)+\frac{2 \pi}{\Lambda}$ is the wavevector mismatch inside the periodically-poled section, $\Lambda\approx 9 \mu m$ is the poling period, $L_{PDC}$ is the length of the PDC section, and $\Omega=1/ \tau$.
The signal and idler wave vectors $k_{s,i}=\frac{n_{s,i}(\omega_{s,i})\omega_{s,i}}{c}$ depend on the refractive indices $n_{s,i}(\omega_{s,i})$ which can be calculated as $n_{s,i}=n_{s,i}^{LiNbO_3}+\delta n_{s,i}$.
The first term can be obtained by using the dispersive Sellmeier formulas \cite{Sellm} for $LiNbO_3$ and the second one appears due to indiffused Titanium atoms and depends on their concentration profile \cite{Strake, param}.
\bibitem{param} For our simulations we use for $\delta n$ parameters from \cite{Strake} and a
width of the waveguide of $7 \mu m$, a titanium-layer thickness of $100 nm$,
and a titanium density of $\rho=5.67*10^{22} cm^{-3}$.
\bibitem{pc} Tunable polarization converters can be realized by putting contacted metallic stripes on the waveguide system and
therefore the rotation angle can be tuned by an externally applied voltage \cite{arxiv, Huang, Moeini}.
\bibitem{parameters}For our numerical solutions we use a rather long pulse of duration $\tau \approx 1 ns$ at a wavelength $780 nm$,
and lenghtes of $L_{PDC}=2.07 cm$, $L_{PBS}=4000 \mu m$, $L_{BS}= 13100 \mu m$, $L_{PC_1}=7620 \mu m$, and $L_{PC_2}=2640 \mu m$.
\end{thebibliography}
\end{document}